\documentclass[12pt]{article}
\usepackage{arxiv}
\usepackage{amssymb}
\usepackage[utf8]{inputenc} 
\usepackage[T1]{fontenc}	
\usepackage{url}        	
\usepackage{booktabs}   	
\usepackage{amsfonts}   	
\usepackage{nicefrac}   	
\usepackage{microtype}  	
\usepackage{graphicx} 	 
\usepackage{subcaption}
\usepackage{doi}
\usepackage{amsmath, calc}
\usepackage{float}
\usepackage{mathabx}
\usepackage{apacite}
\usepackage[toc,page]{appendix}
\usepackage{listings}
\usepackage{xcolor}
\usepackage{setspace} 
\usepackage{natbib}
\usepackage{multicol}
\usepackage{multirow}
\usepackage{array}
\usepackage{makecell}
\usepackage{colortbl}
\usepackage{longtable}

\newcommand{\mx}{{}_{n}m_{x}}
\newcommand{\qx}{{}_{n}q_{x}}
\newcommand{\dx}{{}_{n}d_{x}}

\definecolor{codegreen}{rgb}{0,0.6,0}
\definecolor{codegray}{rgb}{0.5,0.5,0.5}
\definecolor{codepurple}{rgb}{0.58,0,0.82}
\definecolor{backcolour}{rgb}{0.95,0.95,0.92}

\lstdefinestyle{mystyle}{
	backgroundcolor=\color{backcolour},   
	commentstyle=\color{codegreen},
	keywordstyle=\color{magenta},
	numberstyle=\tiny\color{codegray},
	stringstyle=\color{codepurple},
	basicstyle=\ttfamily\footnotesize,
	breakatwhitespace=false,   	 
	breaklines=true,           	 
	captionpos=b,              	 
	keepspaces=true,           	 
	numbers=left,              	 
	numbersep=5pt,            	 
	showspaces=false,          	 
	showstringspaces=false,
	showtabs=false,            	 
	tabsize=2
}

\title{Clustering country-level all-cause mortality data:\\ a review}

\date{}                 	% Or removing it

\renewcommand{\headeright}{}
\renewcommand{\undertitle}{}
\renewcommand{\shorttitle}{A review on clustering country-level mortality data}

\hypersetup{
  pdftitle={A Review on Clustering Mortality Data},
  pdfsubject={stat.AP},
  pdfkeywords={clustering, all-cause mortality, country-level mortality, Human Mortality Database}
}

\author{ {\hspace{1mm}
Pedro M. De Araújo}\\
	School of Mathematics and Statistics\\
	University College Dublin\\
	\texttt{pedro.menezesdearaujo@ucdconnect.ie} \\
	\And
	{Isobel Claire Gormley} \\
	School of Mathematics and Statistics\\
	University College Dublin\\
	\texttt{claire.gormley@ucd.ie} \\
	\And
	{Thomas Brendan Murphy} \\
	School of Mathematics and Statistics\\
	University College Dublin\\
	\texttt{brendan.murphy@ucd.ie} \\
}

\begin{document}
\maketitle

\begin{abstract}

%\textbf{BACKGROUND:}
Mortality data are relevant to demography, public health, and actuarial science. Whilst clustering is increasingly used to explore patterns in such data, no study has reviewed its application to country-level all-cause mortality. %\textbf{OBJECTIVE:} 
This review therefore summarises recent work and addresses key questions: why clustering is used, which mortality data are analysed, which methods are most common, and what main findings emerge. %\textbf{METHODS:}
To address these questions, we examine studies applying clustering to country-level all-cause mortality, focusing on mortality indices, data sources, and methodological choices, and we replicate some approaches using Human Mortality Database (HMD) data. %\textbf{RESULTS:}
Our analysis reveals that clustering is mainly motivated by forecasting and by studying convergence and inequality. Most studies use HMD data from developed countries and rely on k-means, hierarchical, or functional clustering. Main findings include a persistent East-West European division across applications, with clustering generally improving forecast accuracy over single-country models. %\textbf{CONTRIBUTION:}
Overall, this review highlights the methodological range in the literature, summarises clustering results, and identifies gaps, such as the limited evaluation of clustering quality and the underuse of data from countries outside the high-income world.

\end{abstract}

%242/250 words


\newpage

% \tableofcontents

\newpage

\section{Introduction}

The analysis of mortality data plays an important role in demography, public health, and actuarial science \citep{VanRaalte_2021}. Clustering methods have become increasingly popular for studying mortality patterns because they can uncover groups and patterns in the data that might not be immediately obvious. This has led to growing interest in applying these techniques to mortality analysis in recent years.

Despite this growing interest, no prior study has provided a comprehensive review of applications of clustering to all-cause country-level mortality data. This review therefore brings together key information and findings from studies that have employed clustering techniques to analyse such data, hereafter referred to as "mortality data". By doing so, this review aims to encourage further research in this field while highlighting both the main achievements and the remaining gaps in the literature. Specifically, this review addresses the following questions:
    
\begin{itemize}
    \item Why cluster mortality data? The motivations behind clustering analyses in prior work are summarised.
    
	\item What mortality data have been clustered? Interest lies in exploring the mortality indices used, the periods studied, the countries included, and any transformations or data preprocessing techniques employed.

    \item What methods have been used? This review catalogues some of the statistical and computational techniques applied to cluster mortality data, noting which software tools have been utilised. 
    
    \item What are some of the main findings from previous studies? Selected clustering results are shown, highlighting common findings.

\end{itemize}

Regarding the selection of reviewed works, we focus on studies where the majority of the clustered variables are derived from mortality indices present in period life tables. Works that apply clustering methods at any stage of the analysis, even if clustering was not the primary focus, are included. Peer-reviewed papers, conference proceedings and working papers were analysed. The search began with databases such as Web of Science and Google Scholar, followed by an exploration of the references cited in the identified works and studies that cited them. In all, 27 papers were reviewed: 24 peer-reviewed articles, two conference papers and one working paper, starting from 2002, with 2021 being the median year of publication. The list of reviewed papers is given in Appendix \ref{app:papers}.

To illustrate methodological differences, we also replicate some clustering approaches using data from the Human Mortality Database (HMD) \citep{HMD}, focusing on 30 countries and annual period life tables spanning 1960 to 2010. The data were downloaded using the \texttt{HMDHFDplus} package \citep{Riffe_2015}, and the \texttt{R} code with which all results herein were produced is available online.\footnote{\url{https://github.com/pedroaraujo9/review-all-cause-mortality-clustering}}

The remainder of this review is organised as follows: we start by providing definitions of mortality indices in Section \ref{index-def}, with Section \ref{section:motivation} then addressing the primary motivations for clustering mortality data. Section \ref{section:data} discusses the mortality data that have been clustered in previous works, while Section \ref{section:methods} details the clustering methods employed. Section \ref{section:findings} highlights some of the main findings from the reviewed papers. Replication of some clustering methods and their application to a standard HMD dataset are presented in Section \ref{sec:replication}. Finally, a discussion and conclusions are provided in Section \ref{section:conclusion}.

\section{Mortality indices}\label{index-def}

Before discussing the clustering of mortality data, we first outline the main mortality indices used in the reviewed works. The focus is on all-cause country-level mortality data, that is, data without cause-specific detail and aggregated at the national level. Most studies rely on period life tables, which summarise a population’s mortality experience across age groups in a given period for a hypothetical cohort assumed to experience those observed rates throughout their lives. Age groups are typically defined in one-year widths (e.g., $[0,1)$, $[1,2), \dots, [110,+\infty)$) or in five-year widths (starting with $[0,1)$ and $[1,5)$, then $[5,10)$, and so on, ending with an open-ended interval).

Four indices from period life tables are frequently analysed: the death count distribution (Equation (\ref{eq:dx})), the central mortality rate (Equation (\ref{eq:mx})), the probability of death (Equation (\ref{eq:qx})), and the life expectancy (Equation (\ref{eq:ex})). Each index is defined for a specific age interval $[x, x + n)$ for a well-defined population in a well-defined period (in years). For more details on life tables and mortality indices, see \cite{Preston_Heuveline_2000} and \cite{Keyfitz_2013}.

The life table death count is typically defined as
\begin{equation}\label{eq:dx}
_{n}d_{x} = \text{death count between ages } x \text{ and } x+n,
\end{equation}

and represents the number of deaths occurring in a hypothetical cohort. The death count can also be expressed as $_{n}d_{x} = l_{x} - l_{x+n}$, where $l_{x}$ is the number of survivors at age $x$, with $l_{0} = 100{,}000$ for standard life tables. The modal age at death, which is the mode of $_{n}d_{x}$, and the relative death count, $_{n}d^{\star}_{x} =\: _{n}d_{x}/l_{0}$, are quantities of interest derived from the life table death count. 

The life table central mortality rate is defined as
\begin{equation}\label{eq:mx} \mx = \frac{\text{Death count for ages between } x \text{ and } x+n}{\text{Number of person-years lived for ages between } x \text{ and } x+n} =\frac{{}_{n}d_{x}}{{}_{n}L_{x}} \end{equation}

where ${}_{n}L_{x}$ is the total number of person-years lived in the interval, which is the total amount of time (in years) lived by the cohort within the age interval. A population, or empirical, counterpart is often used when constructing period life tables and is given by the ratio of empirical deaths to empirical person-years, namely \({}_{n}M_{x} = {}_{n}D_{x}/{}_{n}E_{x}\). The life table mortality rate and its empirical version usually coincide, although differences may arise at older ages where mortality rates are commonly estimated for life tables \citep{Feehan_2018}. The central mortality rate $\mx$ remains one of the key mortality indices and is the basis of several influential mortality models, including the Lee--Carter model \citep{Lee_Carter_1992}, the Li--Lee ACF model \citep{Li_Lee_2005}, and many related extensions.

Another relevant life table index is the probability of death, defined as
\begin{equation}\label{eq:qx} \qx = \frac{\text{Death count for ages between } x \text{ and } x+n}{\text{Number of survivors to age } x}=\frac{{}_{n}d_{x}}{l_{x}}. \end{equation}

Under some conditions, such as the central mortality rates being constant within the age group $[x, x+n)$, the probability of death $\qx$ is directly connected with $\mx$ by the formula:
$$\qx = \frac{n \cdot \mx}{1 + (n - {}_{n}a_{x})\cdot \mx}, $$

where ${}_{n}a_{x}$ represents the average number of years lived in the interval by those who die between ages $x$ and $x+n$.

Finally, the (remaining) life expectancy, of a person aged $x$, is defined as
\begin{equation}\label{eq:ex} 
e_{x} = \frac{\text{Person-years lived above } x}{\text{Number of survivors to age } x}= \frac{T_{x}}{l_{x}}. 
\end{equation}
The life expectancy at birth $e_{0}$, and the healthy life expectancy at birth \citep{Jagger_2011}, which is the life expectancy at birth adjusted by years lived with disabilities, are related quantities of interest.

This review focuses on studies that mainly use the life table indices or quantities directly derived from these tables to cluster the entities. Some of the reviewed works use only one index for a few or all age groups, while others use a combination of different indices from period life tables with some additional information. Typically, such data are high-dimensional, varying across age groups, countries, periods and sexes, and are highly correlated over time, across age groups and countries.

\section{Why cluster mortality data?}\label{section:motivation}

The primary goal of any cluster analysis is to identify groups (clusters) among the entities being investigated \citep{Everitt_Landau_2010}. Ideally, the entities within a cluster should be as similar as possible to each other and as different as possible to entities in other clusters. When analysing mortality data, these principles remain the same, with the only variation being the purpose of the clustering solution. Broadly, clustering mortality data serves two main purposes: i) improving forecasting or ii) analysing patterns in mortality (such as stages, convergence, and inequalities). Additionally, some studies have used clustering of mortality data to illustrate clustering methodologies, or to estimate life tables.

\subsection{Improving forecasting}
\label{sec:forecasting}

Several clustering studies on mortality data aim to improve forecasting and are inspired by multi-population mortality models (see \cite{Enchev_Kleinow_2016} for a review of these models). For instance, the model proposed by \cite{Li_Lee_2005}, referred to as the augmented common factor (ACF) model, is an extension of the well-known Lee--Carter model \citep{Lee_Carter_1992}, and it produces "coherent" forecasts for multiple populations. For the central mortality rate of age group $x$, period $t$, and population $i$ (suppressing the age group width $n$ subscript prefix), the Lee--Carter model is often fitted to a single population $i$ and is defined as 
$$\log(m_{xit}) = \alpha_{xi} + \beta_{xi}\kappa_{it} + \epsilon_{xit},$$ 
where $\alpha_{xi}$ is the average log mortality for age group $x$ for population $i$, $\beta_{xi}$ is an age group effect for population $i$, $\kappa_{it}$ is a temporal effect for population $i$, which is forecast with a time-series model, usually a random walk with drift, and $\epsilon_{xit}$ is the stochastic error. The ACF of \cite{Li_Lee_2005} extends this Lee--Carter model by adding a common age ($\beta_{x}$) and period ($\kappa_{t}$) effect to consider multiple populations with a "coherent" forecast, i.e.,
$$\log(m_{xit}) = \alpha_{xi} + \beta_{x}\kappa_{t} + \beta_{xi}\kappa_{it} + \epsilon_{xit}.$$
Here "coherent" means there is a constant ratio between the forecast mortalities of different populations, i.e., the ACF model is parametrised such that 
$$\lim_{t\to+\infty}\frac{m_{xit}}{m_{xjt}} = c_{xij} \in \mathbb{R}_{+}, \:\: i\neq j,$$

so the forecasts do not diverge over time.

To improve such multi-population forecasting methods, it is preferable to select similar populations that have similar mortality dynamics and can share common parameters. For this purpose, countries are often the entity being clustered, and the aim is to find groups of countries with similar mortality dynamics across all age groups (or occasionally for a specific age group). The fitting of the required models can be done in independent steps, first clustering the countries and then fitting the forecasting model or the fitting is done in such a way that the clustering and forecast model fitting are done jointly in a single step.

Many works use the two-step approach. For example, \cite{Hatzopoulos_Haberman_2013} first group countries by comparing the time-related parameters extracted from an initial mortality model, and then use the countries from one of the groups to construct a shared age–period model; in \cite{Tang_Shang_2022}, the countries are clustered using a functional panel data model, and then the model is refitted for each cluster to produce forecasts. 

There are two-step applications where the clustering step is not directly related to a mortality model. The works of \cite{Levantesi_Nigri_2022} and \cite{Levantesi_Nigri_2023} cluster countries based on life expectancy quantities, with methods that do not account for any parameters from the forecast model, and then fit a multivariate forecasting model for each cluster. Similarly, \cite{Shen_Yang_2024} first group countries according to similarities in their mortality patterns, and then use these groups to inform the structure of a single model that forecasts mortality for all countries together. Differently from the previous works, \cite{Tsai_Cheng_2021} cluster the age groups (instead of the countries) to look for common age-group effects before fitting a forecast model.

Some researchers cluster and fit the forecasting model in an integrated manner. For example, \cite{Schnrch_Kleinow_2021} propose a common age effect, using a fitting procedure based on fuzzy clustering membership weights, with the clustering and the model being fitted jointly; the authors also explore two other two-step methods based on the Lee--Carter and the ACF model. In \cite{Perla_Scognamiglio_2022}, the authors fit a neural-network model with local coherence, where subsets of countries have coherent forecasts, arguing that a fully coherent model is too strong when modelling many countries jointly. 

\subsection{Analysing patterns} 

Clustering has also been used to investigate patterns in the mortality profiles of countries. Here, the goal is more general, and it includes tasks like identifying signs of mortality convergence between countries, inequality, or overall trends across populations, and possibly explaining what is driving any differences. More specifically, many demographic studies debate whether countries' mortality dynamics follow a common path and converge towards similar levels or not, passing through different mortality stages over time \citep{omran_1971,Wilson_2001,Vallin_Mesle_2004}. There is also interest in understanding inequalities in health systems and outcomes \citep{Gwatkin_2000}, relying on mortality data to describe health across countries.

For some of these tasks, by clustering, for instance, the ``country--year'' or the ``country--year--age'' entities, researchers can track changes in the cluster membership of countries over time, which can help to describe mortality evolution, inequality, and other changing features of the data. In \cite{Mesle_Vallin_2002} countries are clustered in two different periods independently based on the log probability of death to study mortality divergence patterns in Europe. Similarly, \cite{Atance_2024} explore convergence trends by clustering countries based on various mortality indices in three periods independently, the last period being an extrapolation based on a forecast. \cite{Leger_Mazzuco_2021} explore differences in mortality decline by clustering the death count distribution for all ``country--year'' pairs. \cite{Piscopo_Resta_2014} and \cite{Debon_Haberman_2023} cluster the ``country--year--age'' entity to understand how mortality improvements spread across age groups and differ across developed countries.

Many studies cluster solely the countries when looking for patterns in the data. \cite{Debon_Chaves_2017} cluster EU countries and then analyse between-group inequality in mortality (and longevity) using various mortality indicators; \cite{Dimai_2025} investigate convergence in mortality experience by clustering countries based on the death count distribution; \cite{Top_Cinaroglu_2021} cluster European countries based on long-term life expectancy at birth and then study how economic growth and financial development indicators predict these country groups; while \cite{Cefalo_Levantesi_2023} and \cite{Alaimo_Nigri_2024} cluster countries using the life expectancy sex ratio over time to study gender differences in mortality dynamics; finally, \cite{Alaimo_Levantesi_2024} cluster countries based on the age-specific mortality and disability contributions to changes in healthy life expectancy (HLE), in order to summarise and compare HLE dynamics.

Other works cluster countries using a limited time window to investigate current trends. For instance, \cite{Ruger_2006} and \cite{Day_Pearce_2008} cluster countries based on single-year mortality data to investigate global health inequalities; \cite{Zafeiris_2019} characterise and cluster the mortality regimes of Eurozone countries using 2015-2016 data on death rates, longevity measures, and cause-of-death effects.

%and \cite{Agoston_Vaskoevi_2020} cluster European countries to compare mortality experiences.

\subsection{Other reasons}

Another motivation for clustering mortality data is life table construction \citep{Coale_Demeny_1966,UN_1981}. Several model life table systems have been developed to generate complete life tables for populations where vital statistics are missing, unreliable, or inconsistently reported. In such settings, limited observed data for a subset of ages can be combined with an appropriate model to estimate the full age schedule of mortality. Clustering can support this process by identifying groups of populations that share comparable mortality patterns. For example, \cite{Clark_Sharrow_2011} use model-based clustering to group mortality age patterns, and these groups are then used to construct their model life table patterns.

Lastly, some studies use mortality data as an illustrative example to demonstrate the utility of a method without focusing on demographic or actuarial questions. Mortality data are well-suited for this purpose because they represent a multivariate time series at the country level, which can also be treated as multivariate functional data, panel data, or multivariate longitudinal data. The high-quality and public availability of the HMD make it ideal for illustrating various statistical methods. For example, \cite{Fop_Murphy_2018} use the HMD to demonstrate feature selection techniques in clustering. \cite{Lopez-Oriona_2025} also use HMD data to illustrate a novel method to cluster functional time series data. 

\section{What mortality data have been clustered?}\label{section:data}

When clustering mortality data, several choices must be made regarding the data. This section describes the mortality data used in previous studies, introducing the main data sources, commonly used mortality indices, typical transformations, aggregations, data imputations, the number of countries analysed, and other relevant aspects.

\subsection{Data sources}

Regarding multi-country all-cause mortality data sources, the majority of the works reviewed used the HMD, for instance, \cite{Hatzopoulos_Haberman_2013,Schnrch_Kleinow_2021,Debon_Chaves_2017} and \cite{Dimai_2025}. The HMD provides high-quality data for period life tables and other demographic quantities for a select group of countries, mostly developed ones. Appendix~\ref{ap:hmd} provides details on the populations and periods available in the HMD. At the moment, the database includes data from 49 different populations for males, females and combined sexes, with interval sizes for age groups of 1 and 5 years (also called abridged life tables), and periods of 1, 5, and 10 years. The HMD can be loaded directly in \texttt{R} with the package \texttt{HMDHFDplus} \citep{Riffe_2015}, the package \texttt{vital} \citep{Hyndman_2024}, or can be downloaded online.\footnote{\url{https://www.mortality.org/}}

Another mortality database is the World Population Prospects (WPP) mortality dataset \citep{WPP_2024}, used in \cite{Atance_2024}. The WPP database contains period life tables for almost every country and some regional aggregations (296 populations, including countries and regional aggregations), covering data from 1950 onwards yearly for males, females, and combined sexes, and age group intervals of 1 and 5 years. For some developing countries, the WPP database uses estimates for their mortality information, using different strategies, with further details available in \cite{UN_2024}. Finally, the WPP data are available for download online.\footnote{\url{https://population.un.org/wpp/Download/Standard/Mortality/}} 

Mortality data for European countries are also provided by Eurostat\footnote{\url{https://ec.europa.eu/eurostat/statistics-explained/index.php?title=Mortality_and_life_expectancy_statistics}} and used by \cite{Zafeiris_2019}. Healthy life expectancy can be found in the Global Burden of Disease (GBD) Study \citep{GBD_2023},\footnote{\url{https://www.healthdata.org/research-analysis/gbd}} which is used by \cite{Levantesi_Nigri_2023}. Additionally, the World Bank provides mortality indicators for many countries, including life expectancy, which was used by \cite{Top_Cinaroglu_2021}.\footnote{\url{https://data.worldbank.org/}}

\subsection{The set of countries analysed}\label{sec:set-countries}

The set of countries analysed is typically limited to developed countries, given that most studies rely on the HMD data, which offers high-quality data for a limited number of countries. Some works are even more restrictive and focus only on European countries, like \cite{Debon_Chaves_2017,Agoston_Vaskoevi_2020} and \cite{Top_Cinaroglu_2021}, while others, such as \cite{Hatzopoulos_Haberman_2013,Perla_Scognamiglio_2022} and \cite{Dimai_2025}, do not limit their analysis to a single continent. \cite{Day_Pearce_2008,Ruger_2006} and \cite{Atance_2024} are exceptions by applying cluster analysis to many countries.

Excluding countries with a small population or limited data periods is a common practice. For instance, \cite{Tang_Shang_2022}, \cite{Levantesi_Nigri_2022}, among many others, do not consider countries with data not available in the period analysed to ensure comparability, while some works exclude some countries with a small population, such as Luxembourg, which can have more irregular mortality \citep{Leger_Mazzuco_2021,Levantesi_Nigri_2022}. 

\subsection{Choice of mortality index}

The choice of mortality index analysed typically depends on the goal of the research, as cluster analysis is often used as an intermediate step (see Section \ref{section:motivation}). Many studies have employed $\mx$, given that other indices derive from it, and it is commonly modelled for forecasting and other demographic analyses. Examples include \cite{Schnrch_Kleinow_2021, Tang_Shang_2022}, \cite{Perla_Scognamiglio_2022} and many others.

The analysis of life expectancy-derived quantities is also common given that they summarise mortality experience in a single index, and are used for many other purposes, such as being part of indices like the United Nations' Human Development Index (HDI) \citep{HDR}. In \cite{Levantesi_Nigri_2022} and \cite{Levantesi_Nigri_2023}, life expectancy-related indices were clustered for a forecasting application, while in \cite{Cefalo_Levantesi_2023} and \cite{Top_Cinaroglu_2021} they were used to investigate patterns in the data.

The probability of death is considered in \cite{Mesle_Vallin_2002, Debon_Chaves_2017} and  \cite{Agoston_Vaskoevi_2020}, where it is used to investigate patterns in the data. The death count $\dx$ is analysed by \cite{Dimai_2025} and \cite{Leger_Mazzuco_2021}, with the latter stating that it provides a good way to visualise changes in adult population modal mortality.

\subsection{Data transformations}

Although the period central mortality rate $\mx$ theoretically ranges from $[0, +\infty)$ and is not a probability, in practice it is generally observed in the interval $(0, 1)$, especially in modern human life tables, which is the case in the HMD and WPP life tables. Other measures, such as the probability of death $\qx$, are bounded by definition.
For such "bounded" measures, transforming the index to an unbounded space is standard practice. For instance, many studies analyse the log mortality $\log(\mx)$, see \cite{Fop_Murphy_2018,Tang_Shang_2022} and \cite{Perla_Scognamiglio_2022} for example. When working with the probability of death $\qx$, the $\text{logit}(\qx)$ was used by \cite{Debon_Chaves_2017}, who argue that the logit transformation improves goodness of fit and achieves approximate normality and homoscedasticity, making PCA and clustering more robust. Quantities derived from life expectancy $e_{x}$ are usually modelled in their innate form, or their ratio between two different populations is analysed, as in \cite{Alaimo_Nigri_2024}, for example.

Another relevant transformation that is often applied to mortality data is smoothing. Mortality data from 1-year periods or 1-year age groups can be noisy, and smoothing helps reduce random fluctuations so that underlying age patterns become more clearly identifiable. Examples of works that use smoothing procedures include \cite{Tang_Shang_2022} and \cite{Lopez-Oriona_2025}, who smooth $\mx$. Additionally, \cite{Leger_Mazzuco_2021}, \cite{Levantesi_Nigri_2022} and others indirectly smooth yearly mortality data by applying functional clustering methods.

\subsection{Missing data and data imputation}

The main form of missing data, especially in the HMD, is related to data availability, with some countries having shorter time series. As mentioned in Section~\ref{sec:set-countries}, most studies exclude those countries from the analysis. Other databases like the WPP have more standard period sizes across the countries, but many countries have estimated mortality data, given the impossibility of relying on empirical data for all countries.

Other types of missing data are not common in mortality datasets such as the HMD and the WPP database. Death rates of zero, even though they are not missing data, can cause problems when mortality data are transformed to a non-constrained space for analysis using log or logit transformations. Such death rates of zero typically occur in countries with small populations. This issue is not considered particularly concerning, as it is relatively uncommon in the HMD and the WPP databases. Nevertheless, some research has addressed this issue. For instance, \cite{Perla_Scognamiglio_2022} and \cite{Shen_Yang_2024} replace the few zero mortality rates with the average mortality rate across all populations for the corresponding age group, sex, and year. In \cite{Debon_Chaves_2017}, the authors employ a principal component analysis (PCA) method designed to handle missing data \citep{Josse_Husson_2016}.
\cite{Atance_2024}, using the WPP dataset, replace missing values only for the top age group (100+) by using the previous age group in countries where the 100+ data are unavailable.

\subsection{Data aggregation and periods}

Most studies analyse sex-specific data, as males and females tend to exhibit different mortality patterns making sex-specific analysis particularly relevant for demographers and actuaries \citep{Waldron_2005}. Some studies focus on only one sex, such as \cite{Schnrch_Kleinow_2021} (males), \cite{Levantesi_Nigri_2022} (females) and \cite{Debon_Haberman_2023} (males). Many other works, however, analyse both sexes separately, as seen in \cite{Clark_Sharrow_2011,Leger_Mazzuco_2021} and \cite{Dimai_2025}. A smaller set of studies explicitly address a male-to-female (or female-to-male) ratio index, for example \cite{Cefalo_Levantesi_2023} and \cite{Alaimo_Nigri_2024} who model data based on the life expectancy sex ratio. Finally, a few works, such as \cite{Fop_Murphy_2018} and \cite{Shen_Yang_2024}, use combined-sex data only.

With respect to period aggregation, nearly all studies focus on yearly data from the post–World War II era, typically using data from 1950 or 1960 onwards; see, for instance, \cite{Leger_Mazzuco_2021} and \cite{Levantesi_Nigri_2022}. Exceptions include \cite{Day_Pearce_2008, Ruger_2006} and \cite{Zafeiris_2019}, who focus on clustering short periods.

\subsection{Choice of clustering units}

Because of how mortality data are structured, researchers can cluster different entities, such as countries, periods, country–period pairs, age groups, or various combinations of these. The main driver of the choice of entity to be clustered is the motivation for the cluster analysis itself, as discussed in Section \ref{section:motivation}.

Broadly, the literature can be grouped by the type of entity being clustered. Several works cluster countries, either with a focus on forecasting \citep{Hatzopoulos_Haberman_2013,Schnrch_Kleinow_2021,Perla_Scognamiglio_2022} or for pattern analysis \citep{Top_Cinaroglu_2021,Dimai_2025}. Other studies go beyond clustering countries alone and instead aim to explore temporal patterns in more detail. Some cluster countries across two distinct time periods \citep{Mesle_Vallin_2002,Atance_2024}, while others cluster country–period entities across many years \citep{Fop_Murphy_2018,Leger_Mazzuco_2021}. Finally, a smaller set of works cluster age groups \citep{Giordano_Haberman_2019,Tsai_Cheng_2021}, and a few extend the analysis further to age–country–period combinations \cite[e.g.][]{Debon_Haberman_2023}.

\section{Which clustering methods have been employed?}\label{section:methods}

Several clustering techniques have been applied to mortality data. Among the main methods used are established approaches such as hierarchical methods, k-means, fuzzy methods, functional data clustering, and model-based clustering. 

\subsection{Hierarchical methods and k-means clustering}

Hierarchical clustering methods rely on a dissimilarity (or similarity) measure to iteratively agglomerate or divide entities using a chosen linkage function \citep{Everitt_Landau_2010}. Although they require no parametric assumptions, the researcher must still select a dissimilarity metric, linkage function, and criterion for determining the final number of clusters. For example, Ward’s linkage method \citep{Ward_1963} is agglomerative, sequentially merging clusters to minimise the increase in within-cluster variance at each step; it tends to form spherical, equally sized clusters and is sensitive to outliers \citep{Everitt_Landau_2010}.

K-means clustering \citep{MacQueen_1967} and its variants \citep{Ikotun_Ezugwu_2023} are optimisation (or partition) methods requiring the user to predefine the number of clusters, with the algorithm aiming to minimise an objective function. Like Ward's method, k-means minimises within-cluster variance through an iterative procedure: entities are initially assigned to clusters, then repeatedly reallocated based on their distance to a summary measure (e.g., the centroid) until memberships stabilise. Standardising data can change the resulting solution, and, as with Ward's method, k-means tends to form spherical clusters, is sensitive to outliers, and depends on the initial cluster allocation---hence multiple starts are recommended.

Choosing the number of clusters for k-means and hierarchical methods is difficult because the "true" classification is usually unknown and the optimal number depends on the specific data and problem \citep{Hennig_2013}. A good starting point is visualising the data when possible to identify potential cluster shapes. For hierarchical clustering, the dendrogram offers visual hints about successive group formation and cluster distances, though it is only a visual aid. The Elbow Method is widely used: it plots the Within-Cluster Sum of Squares (WSS) against number of clusters $k$, with the ideal $k$ being the point where the reduction in WSS begins to level off (the "elbow"). Additionally, internal quality metrics can support the cluster size choice by measuring within-cluster homogeneity. Many such metrics exist, including the average Silhouette score \citep{Kaufman_1990}, Calinski–Harabasz (CH) \citep{Calinski_1974}, and point biserial correlation (PBC) \citep{Milligan_1985}, each with its own advantages and problems \citep{Mirkin_2011, Schubert_2023}.

Hierarchical methods have been employed many times to cluster mortality data. For instance, \cite{Mesle_Vallin_2002} applied hierarchical clustering after using principal components analysis (PCA) on the probability of death, and then inspected the dendrograms to define geographically coherent groups of countries. \cite{Giordano_Haberman_2019} cluster age groups using Ward’s hierarchical clustering on Euclidean distances to identify homogeneous age-group subpopulations. \cite{Dimai_2025} employed the Hellinger distance on normalised death counts together with complete-linkage, and also used the dendrogram to identify the numbers of clusters. The weighted quadratic deviation (QDEV) proposed by \cite{Arato_Bozso_2009}, was adopted by \cite{Agoston_Vaskoevi_2020} in conjunction with both Ward’s and k-medians algorithms to cluster death probabilities. 

For k-means, \cite{Shen_Yang_2024} used dynamic time warping (DTW) \citep{Giorgino_2009} to measure distances, applying k-means after reducing the dimensionality of $\log(\mx)$ through PCA, selecting the number of clusters based on the Silhouette score and an elbow method. \cite{Schnrch_Kleinow_2021} applied k-means to the age-group effect $\beta_{xi}$ from the Lee--Carter model, selecting the number of clusters based on an information criterion, whereas \cite{Tsai_Cheng_2021} clustered specific age groups directly using k-means and hierarchical methods and selected the number of clusters using several internal quality metrics.

To apply both k-means and hierarchical methods and select the number of clusters based on several internal quality metrics, the \texttt{R} package \texttt{NbClust} \citep{Charrad_Ghazzali_2014} was used by \cite{Tsai_Cheng_2021}. K-means and hierarchical methods are also available natively in R with the \texttt{base} package, and quality metrics can be computed with packages like \texttt{fpc} \citep{Hennig_2024} and \texttt{cluster} \citep{Maechler_Rousseeuw_2025}.

\subsection{Fuzzy methods}

Unlike hierarchical methods and the standard k-means method, fuzzy methods allow the same entity to belong to multiple clusters. This means that each entity will have a membership level $w_g$ for each cluster $g$, where $\sum_{g=1}^{G}w_g = 1$, providing an uncertainty description for the clustering allocation. Among the most popular fuzzy clustering methods are fuzzy c-means \citep{bezdek_1981} and fuzzy c-medoids \citep{Krishnapuram_2001}. Examples of applications of these fuzzy methods include \cite{Hatzopoulos_Haberman_2013}, who used fuzzy c-means to cluster time effects derived from a mortality model and selected the number of clusters by experimenting with different solutions and observing the membership levels. \cite{Debon_Chaves_2017} 
apply PCA to a dataset where each observation is a country and the variables are logit-transformed probabilities of death for multiple ages and years. They retain five principal components and then cluster the countries using fuzzy c-means on these components, using hierarchical clustering as a supporting tool. The number of clusters is chosen by combining information from the dendrogram, the fuzzy membership structure, and cluster validity indices such as the Xie–Beni index \citep{Xie_Beni_1991} and the partition coefficient (PC) \citep{Bezdek_James_1973}.

Additionally, \cite{Alaimo_Levantesi_2024} applied fuzzy c-medoids, and used the Xie--Beni index; \cite{Alaimo_Nigri_2024} implemented a dynamic time warping-based fuzzy c-medoids approach for longitudinal analysis and regular fuzzy c-medoids for cross-sectional analysis, also using the Xie--Beni index and a fuzzy extension for the Silhouette score \citep{Campello_2006}.

The fuzzy c-means method is implemented in the \texttt{R} package \texttt{e1071} \citep{Meyer_Dimitriadou_2023}, and fuzzy c-medoids can be applied using the \texttt{fclust} package \citep{Ferraro_Giordani_2019}, which also includes other fuzzy clustering algorithms and metrics for selecting the number of clusters, such as the Xie-Beni coefficient and the partition coefficient (PC). 

\subsection{Functional methods}

Given the temporal and age-group structure of period life tables, functional methods provide a natural framework for clustering mortality data, since the curves can be expressed as functions. Several functional clustering methods have been developed in the past few years; \cite{Jacques_Preda_2013} and \cite{Zhang_Parnell_2023} provide synoptic reviews.

A popular method to cluster functional data has two steps: a ``filtering" step where the curves are represented by a B-spline basis expansion, and then the coefficients of the basis expansion are clustered with methods such as k-means \citep{Abraham_Cornillon_2003}. The same approach is taken in distance-based functional methods, where a measure of dissimilarity between the functions is computed from their functional representations, and then clustering techniques are applied to the distances \citep{Ferraty_View_2006}. Additionally, there are model-based methods, where the functional representation and the number of clusters are learned simultaneously based on parametric assumptions \citep{James_Sugar_2003,Bouveyron_Jacques_2011}.

Examples of applications of functional approaches to clustering mortality data include \cite{Levantesi_Nigri_2022}, which applied the two-step method to cluster life expectancy at birth for multiple countries, using time as the functional argument and selecting the number of clusters using the elbow method. \cite{Levantesi_Nigri_2023} and \cite{Cefalo_Levantesi_2023} employed a similar approach to \cite{Levantesi_Nigri_2022}. Additionally, \cite{Leger_Mazzuco_2021} applied several functional clustering methods to cluster the age-group death count distribution curves, treating age groups as the function's argument, with each country at a period represented as a different curve.

Regarding the implementation of functional methods, the B-spline coefficients can be calculated with the \texttt{fda} \texttt{R} package \citep{Ramsay_2024}; the model-based approach can be implemented with the \texttt{R} package \texttt{funHDDC} \citep{Schmutz_Jacques_2021} and the package \texttt{fda.usc} can be used to calculate the dissimilarity between functional representations as done in \cite{Leger_Mazzuco_2021}.

\subsection{Forecasting based methods}

As described in Section \ref{sec:forecasting}, many studies cluster mortality data for the purpose of improving forecasts. In a few studies with such a forecasting motivation, the clusters are learned jointly with the parameters of the forecasting model, often selecting the number of clusters based on an information criterion or automatically during model fitting. For instance, in \cite{Schnrch_Kleinow_2021} in one of the explored models, the countries are clustered during the model fitting using ideas from fuzzy clustering methods. \cite{Tang_Shang_2022} propose a functional panel model that jointly estimates the model's parameters and clusters the countries. Additionally, \cite{Perla_Scognamiglio_2022} propose a neural-network method with local coherence where the clusters are learned jointly with the network parameters.

\subsection{Other methods}

Among other methods employed to cluster mortality data, \cite{Piscopo_Resta_2014} applied self-organising maps \citep{Kohonen_2001} to cluster the country--year--age entities.  \cite{Tsai_Cheng_2021} applied Gaussian mixture models alongside k-means and hierarchical methods to cluster age groups using the \texttt{mclust} package \citep{Scrucca_2023}. \cite{Fop_Murphy_2018} applied model-based clustering to log central mortality rates for the country--year entity and compared different variable selection methods; relatedly \cite{Debon_Haberman_2023} applied latent class clustering \citep{Vermunt_Magidson_2002} to cluster the country--year--age entity based on the central mortality rate. Additionally, \cite{Lopez-Oriona_2025} develop fuzzy c-medoids and fuzzy c-means clustering procedures for functional time series, both based on a new dissimilarity metric tailored to functional serial dependence.

\section{What are some of the main findings from previous studies?}\label{section:findings}

When clustering mortality data, the clustering solution is typically a primary output, along with the impact of the clustering solution on the final goal of the analysis. This section summarises the findings from some selected studies in terms of clustering results. As expected, given the diversity of the studies regarding periods, countries, and indices, a full comparison of results is not feasible. Instead, here the goal is to present some of the findings at a high level to explore commonalities in the results.

\subsection{Improving forecasting}

The works of \cite{Tang_Shang_2022} and \cite{Perla_Scognamiglio_2022} focus on improving forecasting accuracy and use the central mortality rate for several age groups, with data starting from 1950 or 1960, sourced from the HMD, and involving a similar number of countries.

\cite{Tang_Shang_2022} cluster countries based on a functional panel data model. They obtain 3 clusters for females, 6 for males and 5 for total (combined-sex) mortality. For females, one cluster groups Eastern European countries such as Belarus, Estonia, Lithuania, Russia and Ukraine together with Iceland; another cluster groups Australia, Canada, New Zealand and the United States; the remaining Western and Northern European countries and Japan form the third cluster. These clusters are then used to fit cluster-specific forecasting models, which produce more accurate forecasts than non-cluster-specific benchmark models.

In \cite{Perla_Scognamiglio_2022}, the clustering and model fitting are done jointly. Even though the data are sex-specific, the country clusters are the same for both sexes, and the model with 4 clusters had the lowest forecasting error. Again, there was a cluster of Eastern European countries with the addition of Portugal; there was a cluster containing countries in Central Europe such as Hungary, Poland, Slovakia, and Czechia as well as the United States and Finland; another cluster contained Western European countries such as Belgium, Denmark, France, the United Kingdom, Sweden and Ireland. The remaining cluster, the largest, comprised other high-income countries such as Australia, Canada, Japan, and others. Compared to other models like the Lee--Carter, the proposed model achieved the overall lowest mean squared error.

\cite{Levantesi_Nigri_2022} applied functional clustering to female life expectancy before fitting a forecasting model. For the period from 1960 to 2003, three clusters were selected. One cluster captured countries with medium longevity including countries such as Ireland, the United States, Finland and others. Another cluster contained low longevity countries such as Bulgaria, Belarus, Ukraine and Russia. A final cluster contained high longevity countries such as Australia, Canada, Switzerland, Iceland, Italy and others.  

We see that across the applications analysed, the East--West division usually emerges, with post-Soviet countries tending to group within the same cluster, and Western developed countries forming another cluster. Nonetheless, the number of clusters tends to go beyond the East--West division, and Central European countries often appear together, while countries such as the United States, Portugal, and Finland occasionally shift between clusters, not always aligning with other Western developed countries. 

\subsection{Analysing patterns}

For applications related to the study of patterns in the data, such as inequality or divergence, it is possible to find works where countries were clustered over different periods. For instance, \cite{Mesle_Vallin_2002} independently cluster the log probability of death of some European countries in 1965 and in 1995 for each gender. A hierarchical method was used, and based on the dendrograms, they conclude that there are four clusters (selected from the male solution): a cluster for Mediterranean and Alpine countries, with countries like Austria, Belgium and Italy; a cluster for the Northern European countries like the United Kingdom, Denmark, and the Netherlands; a cluster with Central European countries like Bulgaria, Czechia and Poland; and a cluster with the former USSR countries, with Estonia, Latvia, Lithuania, Russia and Ukraine. They then compare the four groups in terms of age-specific mortality patterns, and use cause-of-death trends in representative countries to show that East–West differences widened while north–south differences narrowed. They attribute much of this divergence to man-made diseases (such as alcoholism, smoking and accidents) and to the fact that eastern Europe did not benefit from the cardiovascular revolution that reduced mortality in western Europe.

In \cite{Atance_2024}, different mortality indices were clustered based on the WPP data for 1990, 2010, and 2030 (the latter a forecast) to study global convergence and divergence in mortality. Five clusters were identified for each period and sex using multiple clustering quality metrics. In 1990, for example, OECD countries largely formed one cluster, Sub-Saharan countries another, and China grouped with parts of South America, North Africa, and Western Europe. A further cluster included Russia, India, South Africa, and some Polynesian countries, while Rwanda and Uganda stood alone. By 2010, clusters became mostly continent-specific, with exceptions such as Australia, Canada, New Zealand, and Japan joining Western Europe and the United States. The 2030 forecast showed a similar pattern. Overall, all clusters exhibited improvements in life expectancy and smaller sex disparities over time.

Unlike \cite{Mesle_Vallin_2002} and \cite{Atance_2024}, \cite{Leger_Mazzuco_2021} cluster all country–year age-at-death distributions and obtain five clusters per sex, each representing a different mortality profile. The clusters display a clear temporal progression. For males, the early decades are dominated by profiles with high infant mortality, while the Nordic countries show more advanced mortality patterns from the start. One cluster reflects high premature mortality and fewer deaths at older ages, and several Eastern European countries, including Russia, Latvia, Lithuania, Ukraine and Belarus, tend to move into this cluster and remain there. The remaining clusters correspond to increasing concentration of deaths at older ages, with a shift of the modal age toward later life. Most Western countries move over time from early high-infant-mortality clusters to clusters characterised by deaths occurring at older ages.

In summary, the East–West division was relevant again, explored in detail in \cite{Mesle_Vallin_2002} and also appearing in \cite{Leger_Mazzuco_2021}, where Eastern European countries had a distinct clustering allocation over time. In \cite{Atance_2024}, who analyse global data, the Western countries once again form their own cluster across the periods, but the Eastern European countries are not clustered alone this time, being grouped together with some Asian and North African countries, for both males and females. Additionally, sub-Saharan countries are clustered together, while some Latin American countries form another cluster.

\section{Comparing clustering methods using the Human Mortality Database}\label{sec:replication}

In this section, a range of clustering methods are applied to the HMD. We use combined sex data given the illustrative purpose of replicating, where possible, some of the methods described in Section \ref{section:methods}. Data from the HMD, covering 30 countries, are used. Countries with small populations or a limited number of periods were excluded; the final dataset contained yearly periods from 1960 to 2010, resulting in 51 periods. The age groups are spaced in 5-year intervals, starting from $[0, 1)$, then $[1, 5), [5, 10)$, up to $[110, +\infty)$, resulting in 24 age groups. For methods that used the probability of death the last age group was excluded because it is always 1. The dataset has no missing values and no zero death counts; all indices are from period life tables. Here, the focus was on clustering the "country" entity, which was common for both forecasting and analysing patterns applications, as discussed in Section \ref{section:motivation}. 

\subsection{Clustering methods}

The clustering methods considered are based on the hierarchical, k-means, fuzzy, functional, and forecasting-based approaches outlined in Section \ref{section:methods}. We also use the different mortality indices described in Section \ref{index-def}. The specific procedures employed are detailed below, and the code to replicate the analysis is available online.\footnote{\url{https://github.com/pedroaraujo9/review-all-cause-mortality-clustering}}

\begin{itemize}
    \item \textbf{Hellinger--Ward:} Following \cite{Dimai_2025}, we cluster countries using the average Hellinger distance over time of $d_{ixt}^{\star} = d_{ixt}/l_{i0}$. The Hellinger distance between countries $i$ and $j$ over $T$ periods and the set of age groups $\mathcal{X}$ is defined as the average distance between the square roots of the normalised death distributions:
	\[
	D_{ij} = \frac{1}{T} \sum_{t=1}^{T}\sqrt{\frac{1}{2}\sum_{x \in \mathcal{X}} (\sqrt{d_{ixt}^{\star}} - \sqrt{d_{jxt}^{\star}})^2}.
	\]

    We use Ward's linkage method to form the clusters and select the optimal number of clusters based on the dendrogram, Silhouette score, Calinski--Harabasz (CH) and Point Biserial Correlation (PBC) computed with the \texttt{WeightedCluster} package.

    \item \textbf{ILC-k-means:} Inspired by \cite{Schnrch_Kleinow_2021}, the Lee--Carter model is fitted independently for each country using the log mortality rates, and the age-dependent parameter $\beta_{xi}$ is clustered using k-means.
    The Lee--Carter model is fitted with the \texttt{vital} package \citep{Hyndman_2024}, employing a scaled version of the model that reparameterises it so that the drift parameter from the time-series model equals 1. The clustering is performed with the built-in \texttt{kmeans} function in R, and we select the number of clusters based on the Silhouette score, Calinski--Harabasz (CH) and Point Biserial Correlation (PBC), all computed with the package \texttt{WeightedCluster} \citep{Studer_2013}.

    \item \textbf{PCA-fuzzy:} Following \cite{Debon_Chaves_2017}, we consider $\text{logit}(\qx)$, where each row is a country, and each column is a combination of all age groups over the periods, resulting in a $30\times 1174$ matrix, which we standardise. Principal components analysis (PCA) is applied to the data matrix, with the first six principal components, which capture $91\%$ of the original variability, retained. We then cluster the countries using the fuzzy c-means method from the \texttt{e1071} package, with the default setting for the fuzziness parameter $m=2$. The number of clusters is selected by visualising the Silhouette score for fuzzy methods, the partition coefficient, and the Xie--Beni coefficient.

    \item \textbf{Func-k-means:} Following \cite{Leger_Mazzuco_2021}, \cite{Levantesi_Nigri_2022}, and \cite{Abraham_Cornillon_2003}, a two-step functional method is used to cluster life expectancy at birth, $e_{0}$, where the period is treated as the function argument. First, a cubic B-spline with 25 basis functions is fitted independently for each country. The spline coefficients are then clustered using k-means with Euclidean distance, following the same procedure as in the ILC-k-means method.
    
\end{itemize}

Each methodology relies on different mortality indices. As seen in Section \ref{index-def}, the $_{n}d_{x}$, $\mx$, and $\qx$ are mathematically linked, so clustering any of them reflects broadly similar information. The $e_{0}$ summarises the entire mortality curve and is strongly correlated with the underlying age-specific rates, although different mortality profiles can still produce the same $e_{0}$.

The clustering methods, on the other hand, capture different aspects of mortality dynamics. For instance, the ILC-k-means method, by clustering the Lee–Carter $\beta_{x i}$ coefficients across countries, groups countries with similar rates of change in mortality for each age group relative to the overall mortality level $\kappa_{t}$, which can be useful for forecasting. The Hellinger–Ward and PCA–fuzzy methods were used to study inequality and convergence, but they reduce the differences to a single distance measure: Hellinger–Ward does so directly on the observed data, whereas PCA–fuzzy applies the distance in a lower-dimensional representation that preserves the main variability across age groups and periods. Finally, the func-k-means method works with life expectancy at birth, a summary measure of mortality rates, and treats it as a function of time, which can be useful for life-expectancy forecasting or studies concerned with inequality and convergence of life expectancy.

\subsection{Results}

The clustering results for the selected number of clusters under each of the methods considered are displayed in Table \ref{tab:clusters}. Quality metrics are in Appendix \ref{ap:model-selection}.

\renewcommand{\arraystretch}{1.4}
\begin{table}[H]
\centering
\caption{Country cluster composition for each method. For PCA-fuzzy, membership levels for the allocated cluster are shown in parentheses.}
\vspace{0.2cm}
\label{tab:clusters}
\begin{tabular}{p{3cm}p{11cm}}
\toprule
\textbf{Method} & \textbf{Cluster composition} \\
\midrule

\textbf{Hellinger-Ward} &
\textit{Cluster 1:} Australia, Austria, Belgium, Canada, Denmark, Finland, France, Ireland, Italy, Japan, Netherlands, New Zealand, Norway, Portugal, Spain, Sweden, Switzerland, the United Kingdom, the United States.\\
\cmidrule(lr){2-2}
& \textit{Cluster 2:} Belarus, Estonia, Latvia, Lithuania, Russia, Ukraine.\\
\cmidrule(lr){2-2}
& \textit{Cluster 3:} Bulgaria, Czechia, Hungary, Poland, Slovakia.\\
\midrule[1pt]

\textbf{ILC-k-means} &
\textit{Cluster 1:} Australia, Austria, Belgium, Canada, Czechia, Denmark, Estonia, Finland, France, Hungary, Ireland, Italy, Japan, Netherlands, New Zealand, Norway, Poland, Portugal, Slovakia, Spain, Sweden, Switzerland, the United Kingdom, the United States. \\

\cmidrule(lr){2-2}
& \textit{Cluster 2:} Belarus, Bulgaria, Latvia, Lithuania, Russia, Ukraine.\\
\midrule[1pt]

\textbf{PCA-fuzzy} &
\textit{Cluster 1:} Australia (0.91), Austria (0.82), Belgium (0.93), Canada (0.85), Denmark (0.91), Finland (0.79), France (0.91), Ireland (0.82), Italy (0.96), Japan (0.86), Netherlands (0.88), New Zealand (0.85), Norway (0.88), Spain (0.93), Sweden (0.85), Switzerland (0.92), the United Kingdom (0.91), the United States (0.63).\\
\cmidrule(lr){2-2}
& \textit{Cluster 2:} Belarus (0.76), Bulgaria (0.78), Czechia (0.50), Estonia (0.93), Hungary (0.68), Latvia (0.93), Lithuania (0.82), Poland (0.74), Portugal (0.52), Russia (0.86), Slovakia (0.66), Ukraine (0.92).\\
\midrule[1pt]

\textbf{Func-k-means} &
\textit{Cluster 1:} Australia, Austria, Belgium, Canada, Denmark, Finland, France, Ireland, Italy, Japan, Netherlands, New Zealand, Norway, Portugal, Spain, Sweden, Switzerland, the United Kingdom, the United States \\
\cmidrule(lr){2-2}
& \textit{Cluster 2:} Belarus, Bulgaria, Czechia, Estonia, Hungary, Latvia, Lithuania, Poland, Russia, Slovakia, Ukraine.\\
\bottomrule
\end{tabular}
\end{table}

\begin{figure}[H]
\centering
\includegraphics[width=0.8\textwidth]{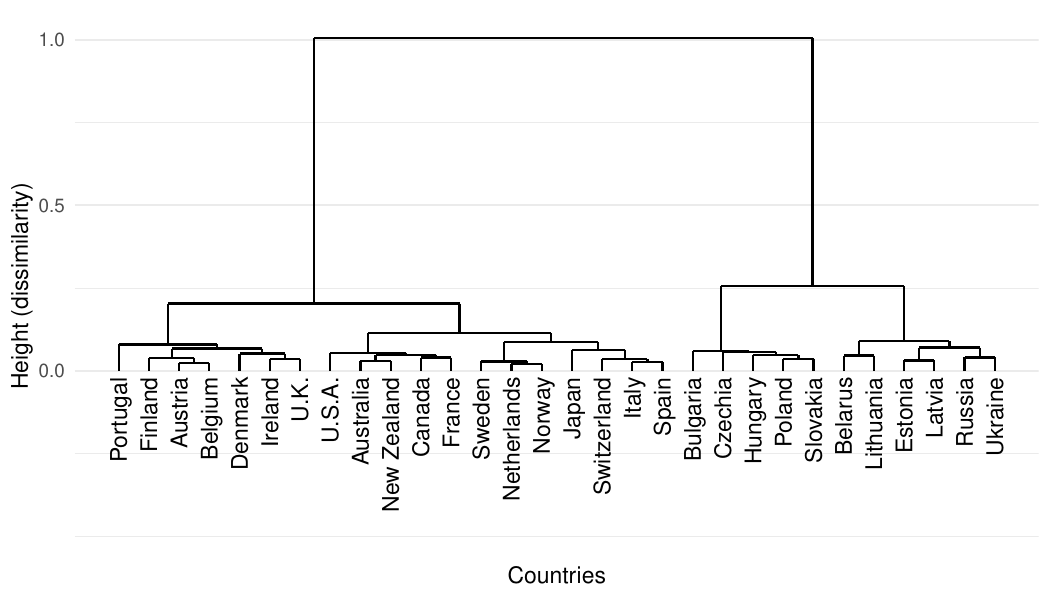}
\caption{Dendrogram for the Hellinger--Ward method.}
\label{fig:dendro}
\end{figure}

The Hellinger--Ward method is hierarchical, allowing us to visualise in Figure~\ref{fig:dendro} the dendrogram representing the successive merges performed by Ward's linkage method. We selected three clusters. The solution with 2 clusters, likewise the other methods, had the best internal metrics, but based on the dendrogram we decided to analyse a solution that provides an extra level of detail. 

 Cluster 1 includes Western developed countries such as Australia, France, Spain, etc. Cluster 2 groups the Baltic countries together with Russia, Belarus and Ukraine. Cluster 3 consists mainly of Central European countries such as Czechia and Poland. For the solution with two clusters, Clusters 2 and 3 are merged, forming the West-East division presented in the other methods.

\begin{figure}[H]
\centering
\includegraphics[width=0.65\textwidth]{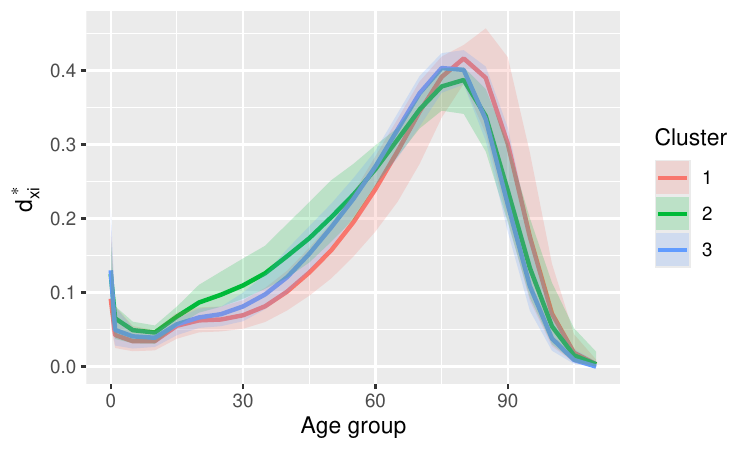}
\caption{Median and 0.025--0.975 quantile bands of the normalised death counts ($d_{xi}^{\star}$) for each cluster.}
\label{fig:dx-cluster}
\end{figure}

Figure \ref{fig:dx-cluster} displays the median normalised death curve ($d_{xi}^{\star}$) for each cluster, along with the 0.025 and 0.975 quantiles. We can see that Cluster 2 (Baltic and Eastern European countries) has a high concentration of adult deaths, followed by Cluster 3 (Central European countries). In Cluster 1 (Western developed countries), the death distribution is more shifted to older age groups than the other clusters.

\begin{figure}[H]
\centering
\includegraphics[width=0.6\textwidth]{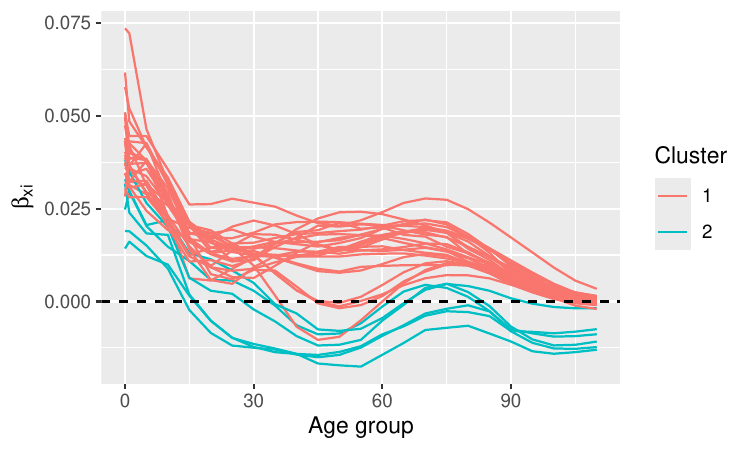}
\caption{Lee--Carter age effect $\beta_{xi}$ for each country coloured by the cluster classification of the ILC-k-means method.}
\label{fig:ILC-k-means}
\end{figure}

For the ILC-k-means we selected two clusters, which presented the highest internal quality metrics. Figure \ref{fig:ILC-k-means} displays the $\beta_{xi}$ parameters of the fitted Lee--Carter model, coloured by the clustering allocation. The ILC-k-means method identified two large clusters: Cluster 1 comprises Western European countries and other developed countries worldwide; while Cluster 2 includes Eastern and Central European countries such as Bulgaria, Hungary, Ukraine, Russia and Estonia. We see that the two groups differ mainly in adult age groups, with some negative values for $\beta_{xi}$ for Cluster 2. 

\begin{figure}[htbp]
\centering
\includegraphics[width=0.80\textwidth]{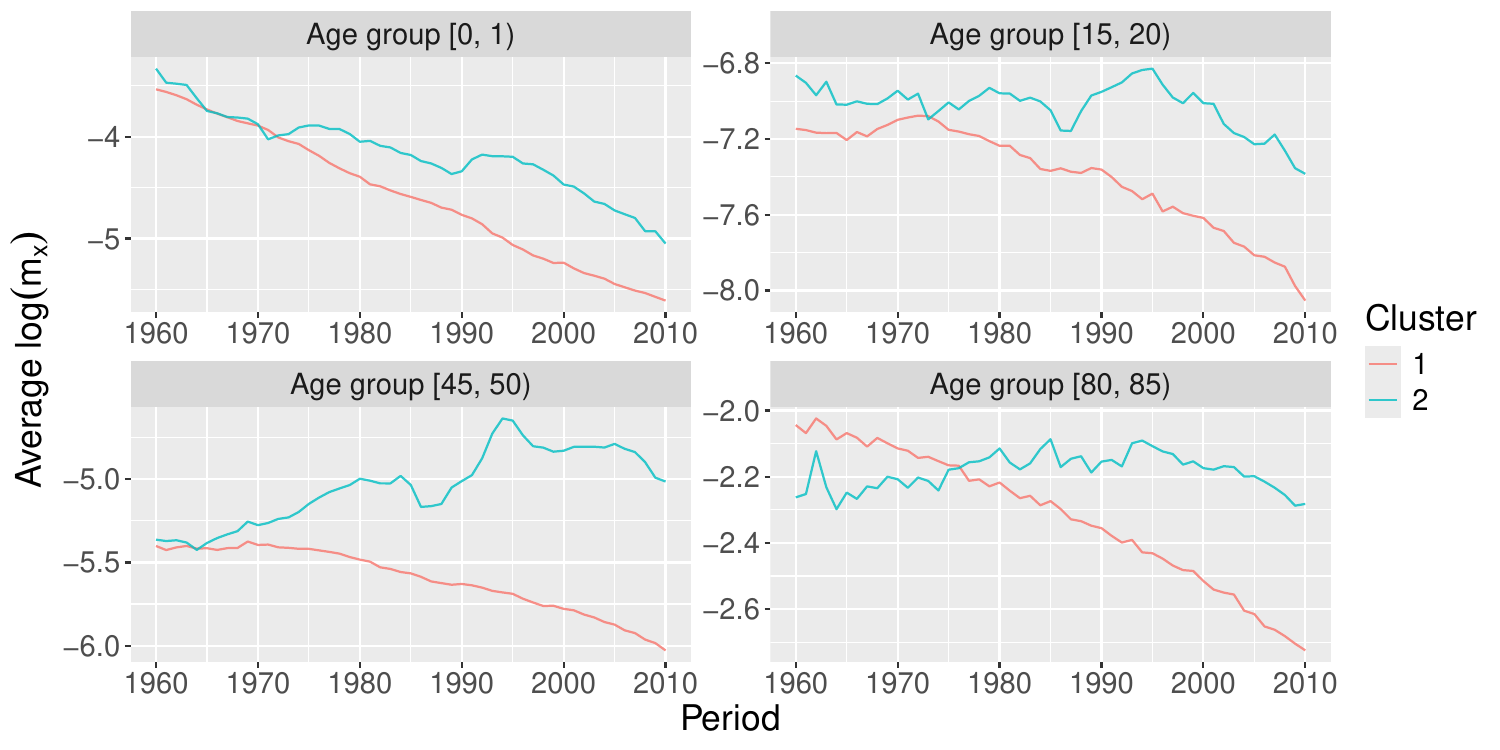}
\caption{Average $\log(_{5}m_{x})$ for four selected age groups for each cluster under ILC-k-means over the periods.}
\label{fig:time-ILC-k-means}
\end{figure}

Figure \ref{fig:time-ILC-k-means} shows the average log mortality for the ILC-k-means clusters over time for four selected age groups. Cluster 1 (Western countries) and Cluster 2 (Eastern European countries) start with a similar average log mortality, but over time the divergence between the clusters increases. For example, for the age group [45, 50), the difference increased from the 1970s until the 1990s, but has been decreasing in more recent periods. Other age groups also present a difference in the average log mortality levels, such as for newborns (age group [0, 1)), where both clusters present a similar decreasing trend in mortality, but Cluster 2 has a higher level and some periods of increasing mortality. 

\begin{figure}[H]
\centering
\includegraphics[width=0.70\textwidth]{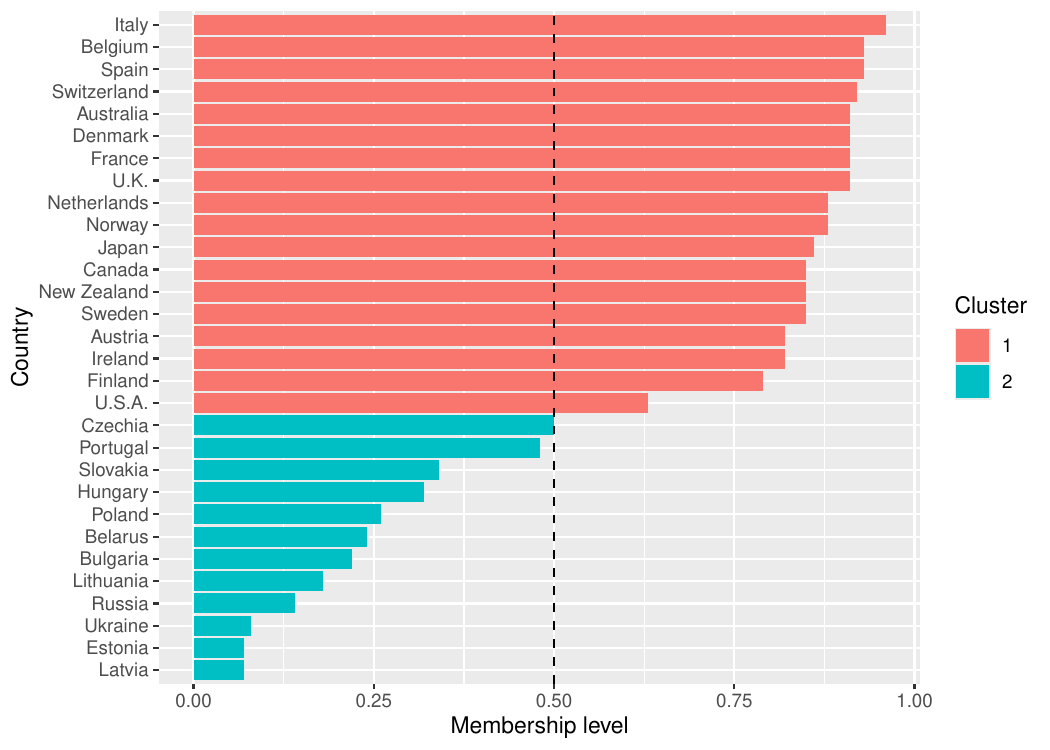}
\caption{Membership level of Cluster 1 (West) for each country for the PCA-fuzzy method.}
\label{fig:fuzzy-level}
\end{figure}

For the PCA-fuzzy method, 2 clusters were selected, as it exhibits the best clustering quality metrics. Table \ref{tab:clusters} details the clustering allocation and again shows the East--West division. Given that we used a fuzzy method, we also have the level of membership for each country, shown in Figure \ref{fig:fuzzy-level}. We see that, for instance, countries like Italy (0.96) and Spain (0.93) have high membership levels for the Western cluster, while countries like the United States (0.63) and Finland (0.79) are the ones with the lowest. For Cluster 2, Portugal (0.52) this time joins the Central and Eastern European countries, with the second lowest membership level for the Western cluster, followed by Czechia (0.50) and other Central European countries.

\begin{figure}[H]
\centering
\includegraphics[width=0.60\textwidth]{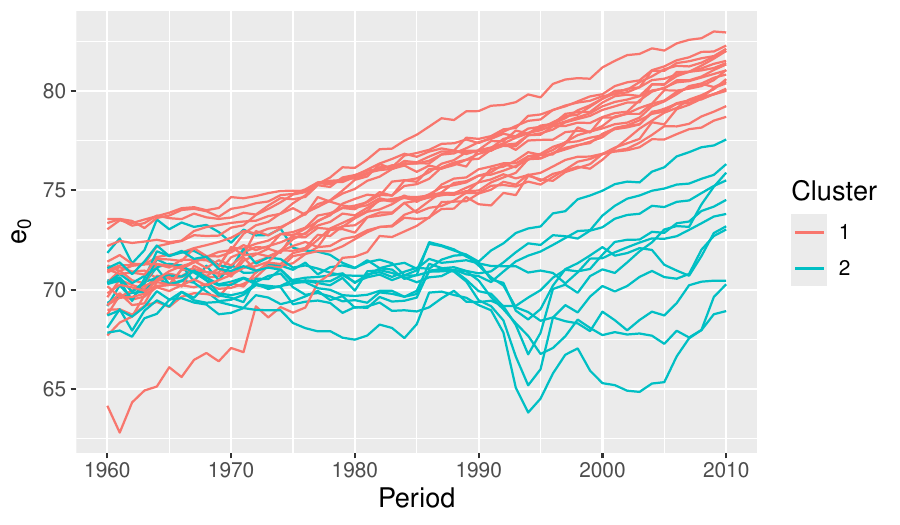}
\caption{Life expectancy at birth over time coloured by clusters of the Func-k-means method.}
\label{fig:func-methods}
\end{figure}

For the Func-k-means once more two clusters were selected based on the best internal quality metrics. The life expectancy at birth coloured by cluster is shown in Figure \ref{fig:func-methods}. Once again, there is separation of developed countries (Cluster 1) and Eastern and Central European countries (Cluster 2). Cluster 2 shows stagnant, and sometimes decreasing, life expectancy for some periods, while Cluster 1 has an increasing life expectancy over the period.

\begin{figure}[H]
\centering
\includegraphics[width=0.80\textwidth]{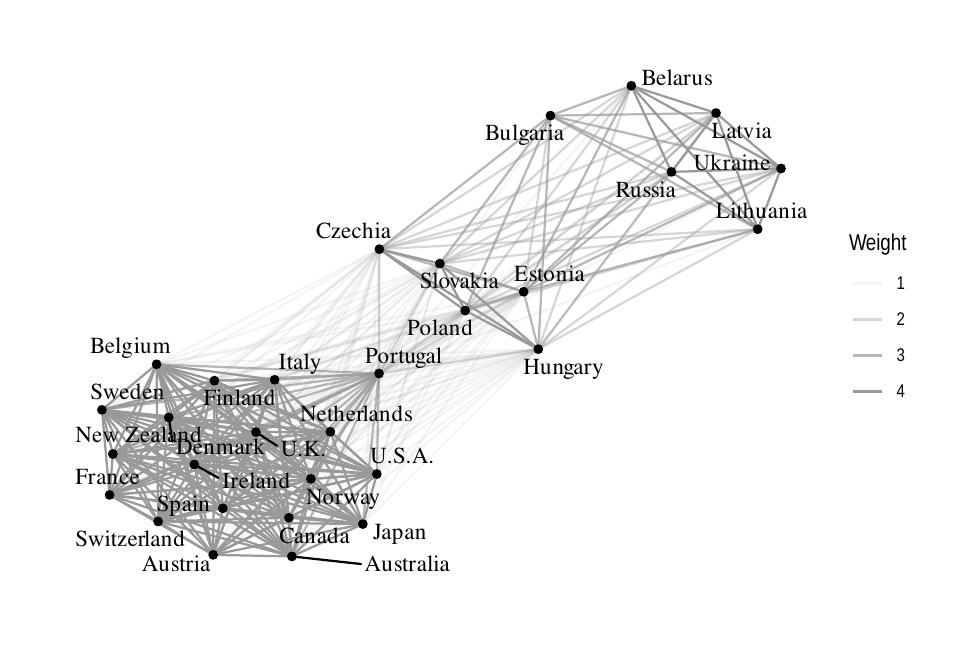}
\caption{The Fruchterman--Reingold layout of the graph where edge weights are the number of times each pair of countries were clustered together across the four methods considered. Countries not clustered together in any of the methods do not have a connection.}
\label{fig:result-graph}
\end{figure}

One way to summarise the information contained in the multiple clustering solutions is to construct a graph where nodes are countries and the weight of an edge between two nodes represents the number of times the countries were clustered together across the methods. Figure \ref{fig:result-graph} shows the Fruchterman--Reingold layout \citep{Fruchterman_Reingold_1991} of such a graph using the clustering solutions given in Table \ref{tab:clusters}. Eastern European countries were most often clustered together, forming a community, while Western European and developed countries formed another community. The Central European countries and Portugal are positioned in the centre of the graph, indicating that they tend to move between the two communities across the methods, with Portugal closer to the Western community, while Czechia, Slovakia, Hungary and Poland are closer to the Eastern European countries across the methods analysed here.

\section{Conclusion and discussion}\label{section:conclusion}

This review was structured around a number of motivating questions. First, why cluster mortality data? Two main motivations were identified: improving forecasting model accuracy, where clustering is usually used as a preparatory step rather than as the primary method, with countries serving as the units being clustered; and analysing patterns in mortality data, such as identifying convergence, inequality, and common trends, where the units can be either countries or country–year pairs, with the latter giving a better description of temporal dynamics. We found relatively few works in other applications, such as model life tables, which suggests potential for further research.

Turning to the data characteristics, a typical clustering application commonly uses the central mortality rate, $\mx$ or $\qx$, across multiple age groups. Another popular choice of mortality index for analysis is life expectancy at birth, $e_{0}$, which simplifies the analysis since each country has only one time series. Data transformations are also common, with $\log(\mx)$ and $\text{logit}(\qx)$ frequently analysed. Additionally, given the high dimensionality of and high correlation in mortality data, it is common to apply a secondary transformation, such as dimensionality reduction methods, typically PCA or smoothing (B-splines). However, the implications of such transformations and dimension reduction methods on mortality data analysis are not well explored in the literature reviewed here and remain an important area for future research.

In terms of data coverage, the data are usually yearly, and mortality indices are typically sex-specific, with clustering results reported separately for males and females. Most studies rely on the Human Mortality Database, which means they focus mainly on developed (largely European) countries and often exclude countries with small populations or limited historical data. Unfortunately, few studies analyse mortality in countries outside the HMD because good empirical data are scarce, particularly for developing countries. While the WPP dataset covers almost all countries in the world and could be used more in applications, it presents complications due to the absence of empirical data for many countries, where model life tables are used instead. Given these data quality concerns, the impact of relying on model life tables when clustering should be carefully investigated before broader adoption.

Regarding methodological choices, many clustering methods have been applied. Popular techniques such as k-means and hierarchical clustering have been widely used, and functional clustering is also frequently implemented, using either time or age group as the functional argument. However, no applications were found that explore the full functional nature of the data across both axes (time and age group), and many applications do not account for time dependence within the data. Recent advancements in time-series clustering and longitudinal data clustering methods suggest these areas are open for further exploration \citep{Lu_2024}.

The investigation of the suitability of model selection techniques, such as the elbow method or internal quality metrics, is rarely addressed, and simulation studies specifically tailored to mortality data remain an area for future research. Such investigations are challenging because true cluster memberships are typically unknown, and the suitability of a method depends on the clustering goal. Nevertheless, in mortality forecasting applications, clustering quality can be directly evaluated through out-of-sample model fit, since the aim is to improve forecasting accuracy, similar to what is done in \cite{Schnrch_Kleinow_2021}. Simulation studies can also highlight the relative advantages of different methods, as demonstrated by \cite{Tang_Shang_2022}.

Regarding empirical results, studies indicate that clustering can enhance forecasting, with models incorporating country groupings generally outperforming those applied to individual countries. A consistent finding across applications is the separation between Eastern and Western European countries in mortality patterns, which is also observed when clustering is used for exploratory analysis. This East--West division was also seen in our replication, although some countries—such as Czechia, Slovakia, and Poland—exhibit cluster memberships that fluctuate between the two groups, with these countries getting closer to the Western pattern in recent periods.

In conclusion, clustering all-cause mortality data at the country level has proven relevant for multi-population forecasting models and for understanding historical patterns, identifying inequalities, and recognising stages in mortality evolution. Although much has been achieved, significant opportunities remain. These include applying recently developed methods that can better summarise and account for the complex dependencies in the data, developing techniques better suited to the particular characteristics of mortality data, exploring more diverse datasets beyond the HMD, and establishing robust frameworks for evaluating the quality of cluster results.

\section*{Acknowledgement}

We thank Dr Ugofilippo Basellini for valuable comments on mortality indices.

This publication has emanated from research conducted with the financial support of Taighde Éireann – Research Ireland under Grant number 18/CRT/6049. For the purpose of Open Access, the author has applied a CC BY public copyright licence to any Author Accepted Manuscript version arising from this submission.

\bibliographystyle{apalike}
\bibliography{references}  

\newpage
\begin{appendices}

\section{List of papers reviewed}\label{app:papers}

\begin{longtable}{@{}l p{9cm}@{}} % two columns: left-aligned reference, left-aligned title
\caption{Reviewed papers (ordered by year of publication) and title.}\label{tab:papers} \\
\arrayrulecolor{black}
\hline
Reference & Title \\
\hline
\endfirsthead

\arrayrulecolor{black}
\hline
Reference & Title \\
\hline
\endhead

\hline
\endfoot

\arrayrulecolor{black}
\hline
\endlastfoot

\cite{Mesle_Vallin_2002} & Mortality in Europe: the divergence between east and west \\

\arrayrulecolor[gray]{0.8}
\hline
\cite{Ruger_2006} & \makecell[l]{Global health inequalities: an international comparison} \\
\hline

\arrayrulecolor[gray]{0.8}
\hline
\cite{Day_Pearce_2008} & \makecell[l]{Twelve worlds: a geo-demographic comparison of global\\ inequalities in mortality} \\
\hline
\cite{Clark_Sharrow_2011} & Contemporary model life tables for developed countries \\ 
\hline
\cite{Hatzopoulos_Haberman_2013} & \makecell[l]{Common mortality modeling and coherent forecasts:\\ an empirical analysis of worldwide mortality data} \\
\hline
\cite{Piscopo_Resta_2014} & \makecell[l]{Multi-country mortality analysis  using self-organising maps} \\
\hline
\cite{Debon_Chaves_2017} & \makecell[l]{Characterisation of between-group inequality of longevity in \\European Union countries} \\ 
\hline
\cite{Fop_Murphy_2018} & \makecell[l]{Variable selection methods for model-based clustering}\\ 
\hline
\cite{Zafeiris_2019} & \makecell[l]{Mortality differentials among the euro-zone countries:\\ an analysis based on the most recent available data} \\
\hline
\cite{Giordano_Haberman_2019} & \makecell[l]{Coherent modeling of mortality patterns for age-specific \\subgroups} \\ 
\hline
\cite{Agoston_Vaskoevi_2020} & \makecell[l]{Clustering EU countries based on death probabilities} \\ 
\hline
\cite{Schnrch_Kleinow_2021} & \makecell[l]{Clustering-based extensions of the common age effect \\multi-population mortality model} \\ 
\hline
\cite{Tsai_Cheng_2021} & \makecell[l]{Incorporating statistical clustering methods into mortality \\models to improve forecasting performances} \\ 
\hline
\cite{Top_Cinaroglu_2021} & \makecell[l]{Cluster analysis of health systems in Europe according to life \\expectancy at birth} \\ 
\hline
\cite{Leger_Mazzuco_2021} & \makecell[l]{What can we learn from the functional clustering of mortality \\data? An application to the Human Mortality Database} \\ 
\hline
\cite{Levantesi_Nigri_2022} & \makecell[l]{Clustering-based simultaneous forecasting of life expectancy \\time series through long-short term memory neural networks} \\ 
\hline
\cite{Tang_Shang_2022} & \makecell[l]{Clustering and forecasting multiple functional time series} \\  
\hline
\cite{Perla_Scognamiglio_2022} & \makecell[l]{Locally-coherent multi-population mortality modelling via \\neural networks} \\ 
\hline
\cite{Levantesi_Nigri_2023} & \makecell[l]{Multi-country clustering-based forecasting of healthy life \\expectancy} \\
\hline
\cite{Cefalo_Levantesi_2023} & \makecell[l]{Modeling gender life expectancy ratio in a multi-population \\framework} \\ 
\hline
\cite{Debon_Haberman_2023} & \makecell[l]{Multi-population mortality analysis: bringing out the \\unobservable with latent clustering} \\ 
\hline
\cite{Alaimo_Nigri_2024} & \makecell[l]{The gender gap in life expectancy and lifespan disparity\\ as social risk indicators for international countries:\\ a fuzzy clustering approach} \\ 
\hline
\cite{Alaimo_Levantesi_2024} & \makecell[l]{Fuzzy clustering of the healthy life expectancy decomposition: \\a multi-population analysis} \\ 
\hline
\cite{Atance_2024} & \makecell[l]{Convergence and divergence in mortality: a global study \\from 1990 to 2030} \\ 
\hline
\cite{Shen_Yang_2024} & \makecell[l]{Advancing mortality rate prediction in European population\\ clusters: integrating deep learning and multiscale analysis} \\
\hline

\cite{Dimai_2025} & \makecell[l]{Clustering of mortality paths with the Hellinger distance \\and visualisation through the DISTATIS technique} \\
\hline

\cite{Lopez-Oriona_2025} & \makecell[l]{Dependence-Based Fuzzy Clustering of Functional Time Series} \\
\arrayrulecolor{black}
\hline
\end{longtable}

\section{Human Mortality Database}\label{ap:hmd}

\begin{figure}[H]
\centering
\includegraphics[scale=0.8]{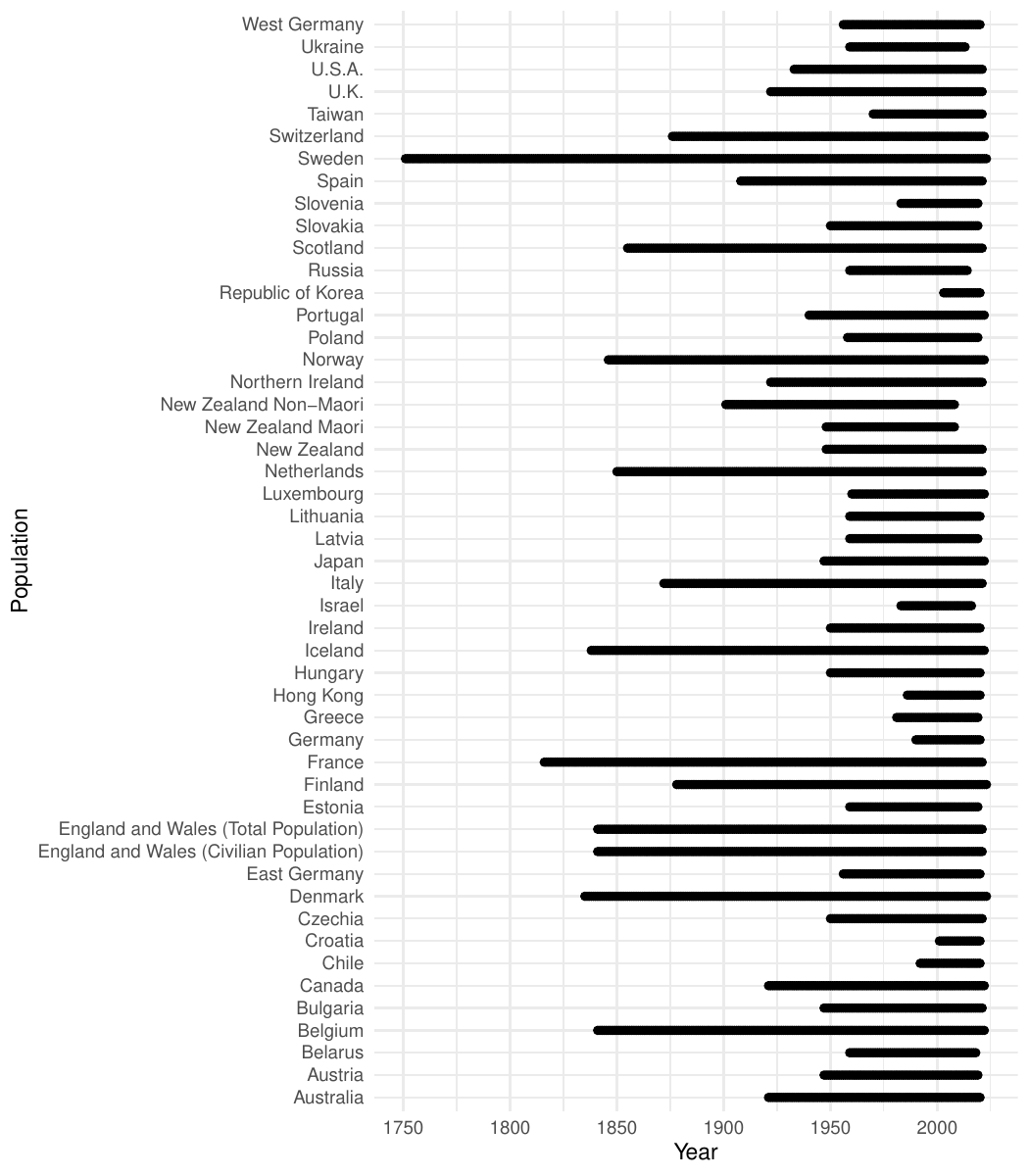}
\caption{Life tables data availability by year for each population on the Human Mortality Database.}
\label{fig:data:data-avail}
\end{figure}

\section{Model selection}\label{ap:model-selection}

As described in \cite{Studer_2013}, the Calinski--Harabasz index (CH) is a variance-based criterion ranging from $0$ to $+\infty$, where higher values indicate better clustering. It evaluates cluster homogeneity by favouring solutions with low within-cluster variance and high between-cluster variance. The Point--Biserial Correlation (PBC) ranges from $-1$ to $1$ (with higher values being better) and measures how well the clustering reproduces the observed pairwise distances. The Silhouette score (or average Silhouette width) also ranges from $-1$ to $1$ (higher being better) and assesses the consistency of the cluster assignments, favouring large between-cluster distances and strong within-cluster cohesion.

For the fuzzy clustering methods, the Partition Coefficient (PC) ranges from $1/G$ to $1$ (higher being better) and measures the fuzziness of the partition, favouring solutions with clearer cluster memberships. The Fuzzy Silhouette score generalizes the classical Silhouette score to account for the degree of membership of each observation. The Xie--Beni index ranges from $0$ to $+\infty$ (lower being better) and evaluates both the compactness and separation of the clusters, favouring solutions with low within-cluster distances and high between-cluster distances.

\begin{figure}[H]
\centering
\includegraphics[width=0.92\textwidth]{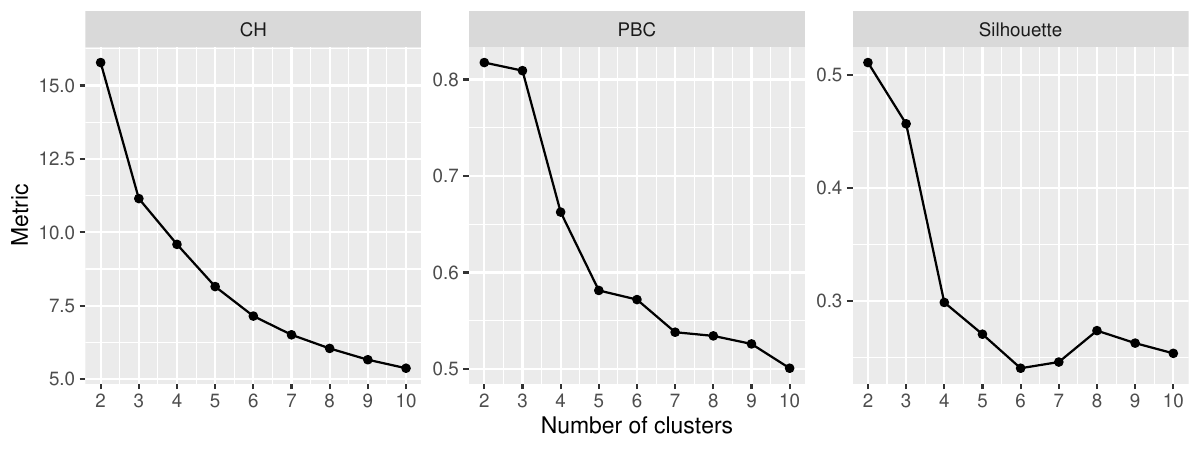}
\caption{Calinski-Harabasz (CH), point biserial correlation (PBC) and Silhouette score for the Hellinger-Ward method.}
\label{fig:hell-ward-metrics}
\end{figure}

\begin{figure}[H]
\centering
\includegraphics[width=0.92\textwidth]{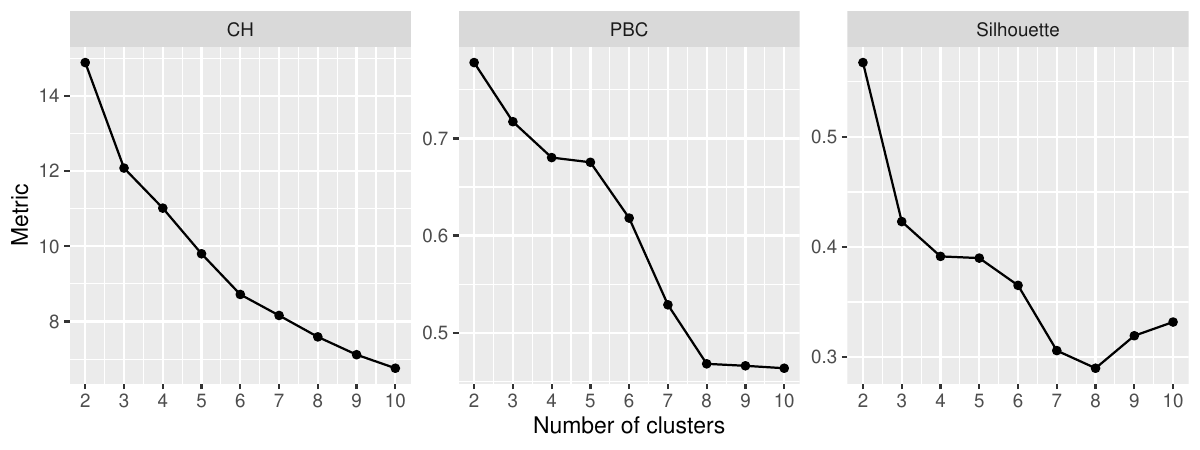}
\caption{Calinski-Harabasz (CH), point biserial correlation (PBC) and Silhouette score for the ILC-k-means method.}
\label{fig:ilc-kmeans-metrics}
\end{figure}

\begin{figure}[H]
\centering
\includegraphics[width=0.92\textwidth]{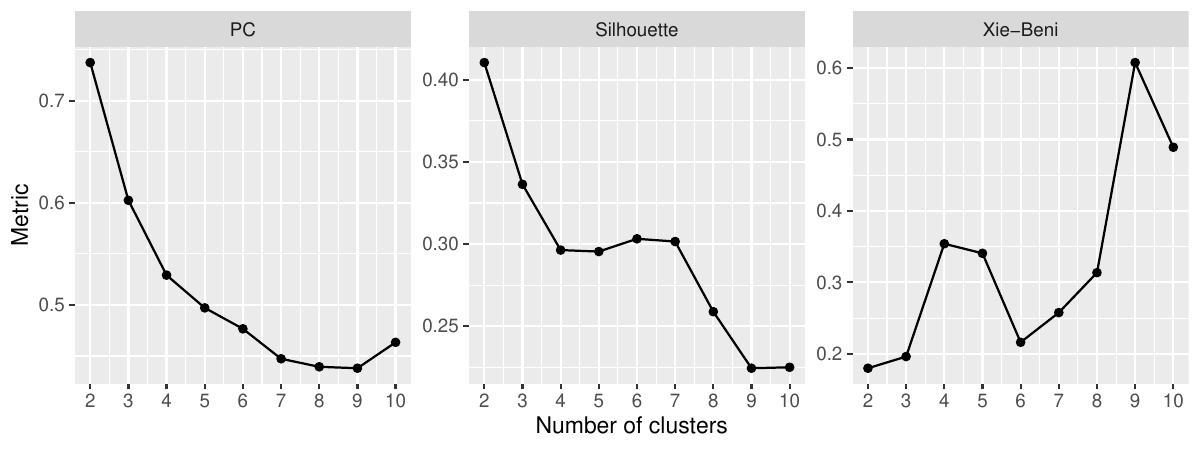}
\caption{Partition coefficient (PC), Silhouette score, and the Xie-Beni coefficient for different numbers of clusters for the PCA-fuzzy method.}
\label{fig:pca-fuzzy}
\end{figure}

\begin{figure}[H]
\centering
\includegraphics[width=0.92\textwidth]{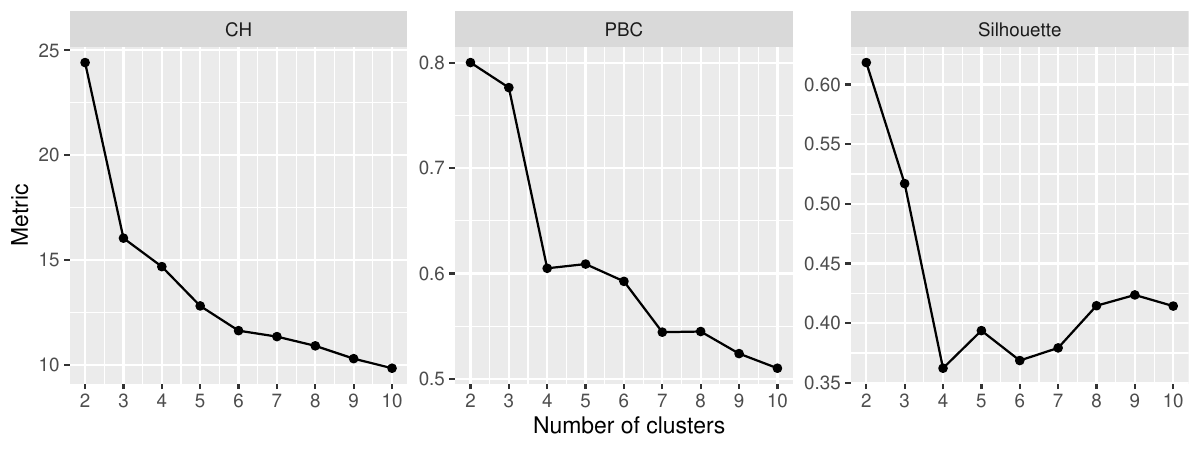}
\caption{Calinski-Harabasz (CH), point biserial correlation (PBC) and Silhouette score for the func-k-means method.}
\label{fig:func-kmeans-metrics}
\end{figure}

\end{appendices}

\end{document}